\documentclass[aps,prb,preprint,12pt]{revtex4-1}

\usepackage{amsmath,amssymb,latexsym}
\usepackage{euscript}

\usepackage{graphicx}
\usepackage{tikz}
\usepackage{pgfplots}

\usepackage[normalem]{ulem} 
\usepackage{soul}           
\usepackage{xcolor}

\usepackage[makeroom]{cancel} 

\usepackage{hyperref}

\begin{document}

\title{What is the maximum radius of cold planets?} 
\author{David Garfinkle}
\email{garfinkl@oakland.edu}
\author{Alberto G. Rojo}
\email{rojo@oakland.edu}
\affiliation{Department of Physics, Oakland University, Rochester, MI 48309}



\begin{abstract}
Planets have maximum radii close to that of Jupiter.  Qualitatively, the reason for this maximum size is that, as one adds mass,  the force of gravity becomes sufficiently strong to cause the radius to decrease.  We show that this effect can be understood quantitatively  using a simple variational principle very similar to that used to compute the size of the hydrogen atom. 
\end{abstract}

\maketitle

\section{Introduction}

This paper presents a simplified variational approach to explain the upper size limits of cold planets and stars,  where by ``cold'' we mean those without significant energy sources to counteract gravitational compression. 
Our approach is based on order-of-magnitude estimates and can be used in undergraduate courses.

Imagine building a material from atoms. The material grows larger, going from the size of a dust grain to that of a lentil and then to that of a small rock. 
If we double the number of atoms, the volume also doubles.
But each increase in the number of atoms represents an increase in gravitational force, and there comes a critical point  where the gravitational force is so great that, if we keep adding atoms, the size starts to decrease instead of increase. 
We explore the dependence of this maximum size on fundamental gravitational and electrostatic constants.

\section{Variational treatment}
\label{VarTreat}
We consider a planet composed of hydrogen atoms under the effect of gravity and seek to determine its maximum size.
The hydrogen atom has a wave function (up to normalization) of $\psi_0 = {e^{-r/{a_0}}}$, where $a_0$ is the Bohr radius. 
The corresponding energy is ${E_0}= -{e^2}/(8\pi \epsilon_0{a_0})$,
 which is the lowest eigenvalue of Schr\"odinger's equation $\mathcal{H} \psi_0(r)=E_i\psi_0(r)$, with $\mathcal H$ the hydrogen atom Hamiltonian.
Note that
 we use $e$
 to denote both the elementary charge and Euler’s number 
$e\simeq
2.718
$. While this might seem ambiguous, it follows standard conventions in physics textbooks.
 
The variational principle states that for any trial wave function \( \psi_{\text{trial}} \), the expectation value of the Hamiltonian
\begin{equation}\langle \mathcal H \rangle={  \int dv \,\psi^*_{\text{trial}}(r) \mathcal H  \psi_{\text{trial}}(r) \over{\int dv\, \psi^*_{\text{trial}}(r)  \psi_{\text{trial}}(r)} }
\end{equation}
provides an upper bound to the true ground state energy \( E_0 \), that is, \( E_0 \leq \langle \mathcal H \rangle \). By optimizing \( \psi_{\text{trial}} \) to minimize \( \langle \mathcal H \rangle \), one can obtain the best approximation for \( E_0 \) from  the  trial function. 

The Hamiltonian for the hydrogen atom is
\begin{equation}
  \mathcal  H = - {\frac {\hbar^2} {2m}} {\nabla^2} - \frac {e^2}{4\pi \epsilon _0 r},
\end{equation}
where $\hbar=h/(2\pi)$ is the reduced Planck constant and $m$ is the reduced mass of the electron.

If our trial wave-function is $\psi_{\text{trial}}={e^{-r/a}}$
where $a$ is some number different from $a_0$, then a standard calculation shows that 
\begin{equation}
E \equiv  \langle \mathcal H \rangle ={\frac {\hbar ^2}{2m{a^2}}} - {\frac {e^2} {4\pi \epsilon _0 a}}.
\label{Ehydrogen}
\end{equation}

The Bohr radius can be derived from this expression by asking what value of the variational parameter $a$ minimizes the energy $E$.  The result, after setting 
\begin{equation}
    {\partial E\over \partial a}=0,
\end{equation}  is exactly
$a={4\pi \epsilon_0\hbar ^2}/(m {e^2})={a_0}$.  We will find that, once the problem of planet size is set up in an analogous  way, the calculation of planet size will be remarkably similar to that of the Bohr radius.

Our first step is to develop a toy model for hydrogen under extreme pressure.  We assume that with $N$ hydrogen atoms in a volume $V$, the nuclei are distributed in a regular cubic lattice with spacing $2a$, and that the total energy is just $N$ times the energy given in Equation (\ref{Ehydrogen}).  In other words, we treat the atoms as non-interacting, with the only effect of the compression being to change the scale in the wave function. 
While there is no reason to expect this  model to be a rigorous representation of the wave function of this collection of electrons, it can still provide insight into how the energy scales with pressure.

Now,  the assumption about the lattice spacing means that 
\begin{equation}
a = {\frac 1 2}{{\left ({\frac V N}\right )}^{1/3}} = {\frac 1 2}{{\left ({\frac {V {m_H}} M}\right )}^{1/3}},
\end{equation}
where $M=Nm_H$ is the mass of the planet and $m_H$ is the mass of a hydrogen atom.  Let $R$ be the radius of the planet.  Then since $V=(4\pi/3){R^3}$ we have
\begin{equation}
    R = a {{\left ( {\frac {6N} \pi} \right ) }^{1/3}}.
    \label{Rfroma}
\end{equation}

For simplicity, we assume that the planet has a uniform density.   Then its gravitational potential energy is
\begin{equation}
{E_{\rm grav}} = - {\frac 3 5} {\frac {G {M^2}} R}.
\end{equation}

We find it useful to express our result in terms of a dimensionless number  $N_*$ that expresses the relative weakness of gravity compared to electromagnetism, as follows:

\begin{equation}
    N_*\equiv\left({e^2\over 4\pi \epsilon_0 Gm_H^2}\right)^{3/2}=\left(1.23 \times 10^{36}\right)^{3/2}=1.37 \times 10^{54}.
    \label{Nstardef}
\end{equation}
In physical terms, $N_*^{2/3}$ is the ratio of electric repulsion to gravitational attraction between two protons. 

The gravitational energy can therefore be written in terms of $N_*$ as 
\begin{equation}
{E_{\rm grav}} = - {\frac 3 5} {\frac {G {(Nm_H)^2}} a \left(\pi\over 6N\right)^{1/3}}=-
N {\frac 3 5} {e^2\over a}{{Gm_H^2 \over e^2}\left(\pi N^2\over 6\right)^{1/3}} =- N {\frac 3 5} {e^2\over 4\pi \epsilon_0 a}\left({\pi N^2\over 6N_*^2}\right)^{1/3}.
\end{equation}

The total energy then takes the form
\begin{equation}
E = N \left [ {\frac {\hbar ^2} {2m{a^2}}} - {\frac {{\tilde e}^2}{ 4\pi \epsilon_0 a}} \right ],
\label{planetenergy}
\end{equation}
where the quantity ${\tilde e}^2 $ is given by
\begin{equation}
{{\tilde e}^2} = {e^2} \left [ 1 + {\frac 3 5} {{\left ( {\frac {\pi {N^2}} {6 {N_*^2}}} \right ) }^{1/3}} \right ]. 
\label{ModifCharge}
\end{equation}

We can think of $\tilde e$ as an $N$-dependent ``modified charge'' that takes into account the effects of both electrostatics and gravity. Notice that the dependence on $N$ becomes relevant for large particle number, $N\sim N^*$.

The planet will adjust its radius (which is equivalent to adjusting the lattice spacing) so as to have the smallest total energy, so we  differentiate Eq. (\ref{planetenergy}) with respect to $a$, set the resulting expression to zero, and solve for $a$; this gives
\begin{equation}
    a= {4\pi \epsilon_0\frac {\hbar ^2} {m {{\tilde e}^2}}} = {\frac {a_0} {1 + \displaystyle{} {\frac 3 5} {{\left ( {\frac {\pi {N^2}} {6 {N_*^2}}} \right ) }^{1/3}}}}.
\end{equation}

Hence, the lattice spacing is given by a  ``modified Bohr radius," with a charge enhanced according to Equation (\ref{ModifCharge}).   Using Eq. (\ref{Rfroma}), we obtain the radius as
\begin{equation}
R (N)=  {\frac {{a_0} {{\left ( \displaystyle{\frac {6N} \pi} \right ) }^{1/3}}} {1 + \displaystyle{\frac 3 5} {{\left ( {\frac {\pi {N^2}} {6 {N_*^2}}} \right ) }^{1/3}}}}.
\label{planetradius}
\end{equation}
When $N\ll N_*$ we have $R\propto N^{1/3}$, and volume scales linearly with the number of particles. On the other hand, when $N\gg N_*$ we have 
$R\propto N^{-1/3}$ and the planet size decreases with particle number.

The number of atoms that results in the maximum planet size can be found from Eq.~(\ref{planetradius}) {\color{black} by finding the value of $N$ that maximizes $R(N)$.  We also use the fact that the total mass $M$  is given by $M=N{m_H}$ to find $M$.  The results are
\begin{eqnarray} 
{R_{\rm max-size}} &=& {\sqrt {\frac 5 {2 \pi}}} {N_* ^{1/3}} {a_0},
\label{MaxRadius}
\\
{M_{\rm max-size}} &=& {\frac 5 3} {\sqrt {\frac {10} \pi}} {N_*} {m_H}.
\label{MassAtMax}
\end{eqnarray}

Note that $N_*$ is independent of $\hbar$, which means that the mass at which maximum planet size occurs is independent of $\hbar$. The maximum planet size, on the other hand, depends on $\hbar$  through the dependence of the Bohr radius on $\hbar$.  

When we substitute the numbers into eqns. (\ref{MassAtMax}) and (\ref{MaxRadius}), we find 
\begin{eqnarray}
{M_{\rm max-size}} &=& 6.86 \times {{10}^{27}} \, {\rm kg}= 3.45\times 10^{-3}M_\odot = 3.6 M_J
\\
{R_{\rm max-size}} &=& 5.2 \times {{10}^7} \, {\rm m}=0.076 \,R_\odot
=0.75 R_J
\label{Rmax}
\end{eqnarray}
Here $M_J=1.90 \times {{10}^{27}} \, {\rm kg}$ is the mass of Jupiter and $R_J=7.14 \times {{10}^7} \, {\rm m}$ is the radius of Jupiter.  More sophisticated calculations 
yield a maximum planet size of ${R_{\rm max-size}}=1.165 {R_J}$  with a mass at maximum size of 
 ${M_{\rm max-size}}=3.31{M_J} $ for a planet made of hydrogen atoms\cite{Zaplosky,Fortney} , and
 a maximum radius of about $1.2 R_J$ and a mass at maximum size of about $3M_J$ 
 for the correct composition of gas giants. Gas giant planets have a chemical composition of 3/4 hydrogen and 1/4 helium (in mass proportions), which is roughly the composition of Jupiter and the Sun. While forming, these planets grew massive enough to accrete all available material along their orbits.  For the purposes of our estimation, one of the crude approximations that we are making is a composition of pure hydrogen. So these numbers are surprisingly good for the size and the mass.

\section*{Conclusions}

We have developed a simplified variational approach to estimate the maximum radius of cold planets. The  radius of such planets results from a balance between electrical attraction, gravitational compression, and  opposing quantum mechanical effects driven by the uncertainty and exclusion principles. 
Although we have assumed a pure-hydrogen planet, our results are in reasonable accord with more detailed models.

 It is important to bear in mind the limited class of  objects to which our approach can be applied.  We consider only ``cold'' objects: that is, ones with no sources of {\color{black}{radiative}} energy ({\color{black} {\em {i.e.}} there is no fusion in the core or external radiation causing atmospheric inflation)}.  The only stars to which our approach applies are those in which fusion has ceased, that is, (1) white dwarfs which are not hot enough to fuse the carbon and oxygen of which they are made, or (2) the iron cores of very massive stars (because iron is the most stable element is the last stage of exothermic fusion).

This  treatment  illustrates the utility of basic physics principles in astrophysical contexts and provides students and researchers with an accessible analytical tool to understand the fundamental limits of planetary and stellar structure.

DG was supported by NSF grant PHY-2102914 to Oakland University. We thank Alejandro Garc\'ia and Alfredo Caro for useful comments on the manuscript.

\end{document}